\documentclass[twocolumn,showpacs,preprintnumbers,amsmath,amssymb]{revtex4}
\usepackage{graphicx}
\usepackage{dcolumn}
\usepackage{bm}
\usepackage{natbib}
\begin{document}
\preprint{Published in \textit{Nature}, \textbf{437} pp. 1330-3 (27
Oct. 2005)}
\title{Observation of spin Coulomb drag in a \\two-dimensional electron gas}
\author{C.P. Weber}
\email{cpweber@lbl.gov}
\author{N. Gedik}
\author{J.E. Moore}
\author{J. Orenstein}
\affiliation{Physics Department, University of California, Berkeley
and \\ Materials Science
Division, Lawrence Berkeley National Laboratory, Berkeley, CA 94720}%
\author{Jason Stephens}
\author{D.D. Awschalom}
\affiliation{Center for Spintronics and Quantum Computation, University of California, Santa Barbara, California 93106, USA}%
\pacs{42.65.Hw, 72.25.b, 74.25.Gz, 75.40.Gb., 78.47.+p}

\begin{abstract}An electron propagating through a solid carries spin angular
momentum in addition to its mass and charge. Of late there has been
considerable interest in developing electronic devices based on the
transport of spin, which offer potential advantages in dissipation,
size, and speed over charge-based devices\protect{\cite{AwschBook}}.
However, these advantages bring with them additional complexity.
Because each electron carries a single, fixed value ($-e$) of
charge, the electrical current carried by a gas of electrons is
simply proportional to its total momentum. A fundamental consequence
is that the charge current is not affected by interactions that
conserve total momentum, notably collisions among the electrons
themselves\protect{\cite{Ziman}}. In contrast, the electron's spin
along a given spatial direction can take on two values, $\pm
\hbar/2$ (conventionally $\uparrow, \downarrow$), so that the spin
current and momentum need not be proportional. Although the
transport of spin polarization is not protected by momentum
conservation, it has been widely assumed that, like the charge
current, spin current is unaffected by electron-electron (\textit{
e-e }) interactions. Here we demonstrate experimentally not only
that this assumption is invalid, but that over a broad range of
temperature and electron density, the flow of spin polarization in a
two-dimensional gas of electrons is controlled by the rate of
\textit{e-e} collisions.
\end{abstract}

\maketitle

In this work spin diffusion is characterized by the transient spin
grating technique\protect{\cite{Cameron}}, which is based on optical
injection of spin-polarized electrons. The two-dimensional electron
gas (2DEG) resides in a GaAs quantum well, in which the carriers are
donated by Si impurites doped into the GaAlAs barrier layers.
Near-bandgap illumination of the GaAs excites electrons whose
initial spin is determined by the helicity of the
light\protect{\cite{OpticalOrientation}}. If the GaAs is excited by
two non-collinear, coherent beams of light with orthogonal linear
polarization, then in the region where the beams interfere the
helicity varies sinusoidally from plus to minus one. The
optical-helicity wave generates a wave of electron-spin polarization
with the same spatial frequency, which in turn generates a
sinusoidal variation (grating) in the index of refraction through
the Kerr effect. The wavevector of the injected spin-density wave is
in the plane of the 2DEG and the spin polarization is oriented
perpendicular to this plane.

The time-evolution of the transient spin grating directly reveals
the nature of spin transport and relaxation in the electronic
system, functioning like a time-domain version of neutron
scattering. We measure the spin polarization by detecting the
diffraction of a probe beam off the grating. A sensitive coherent
detection scheme (described under Methods) enabled acquisition of
the $\sim$ 150 grating decays required to characterize the spin
dynamics for each sample throughout the temperature-wavevector
($T-q$) parameter space.

In this paper we present results for three quantum well samples,
with electron concentrations of 7.8, 4.3, and 1.9 $\times 10^{11}$
cm$^{-2}$, corresponding to Fermi temperatures of 400, 220, and 100
K, respectively. Fig. 1 shows the initial decay rate of the spin
grating as function of $T$ in the most heavily doped sample, for
several grating wavevectors from 0.4$\times 10^4$ cm$^{-1}$ to
2.5$\times 10^4$ cm$^{-1}$. The dependence on $T$ can be described
in terms of three regions. For 100 K $<T<$ 300 K the decay rate
varies slowly. For 50 K $<T<$ 100 K the decay rate increases rapidly
with decreasing $T$, and for $T<$ 50 K it reaches a slowly varying
plateau.

\begin{figure}
\protect{\includegraphics[width=3in]{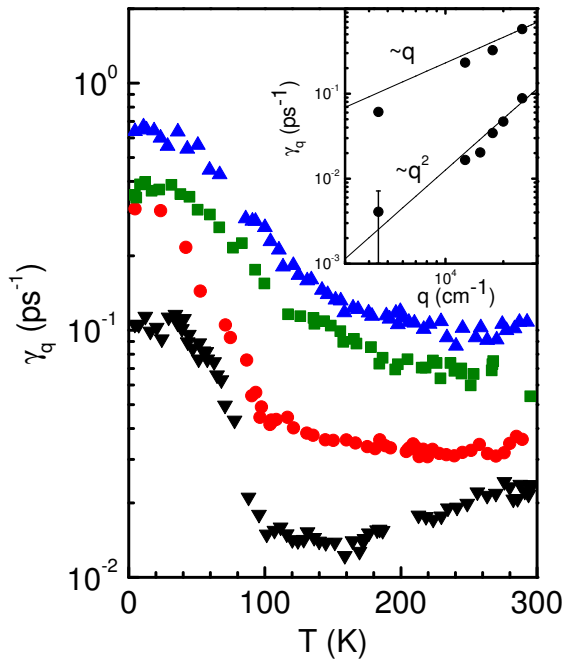}}
\caption{Spin-grating decay at various $q$, $T$ for the sample with
$T_F=$ 400 K. \textbf{Main panel:} The initial decay rate,
$\gamma_q$, of the spin grating as a function of $T$ for (bottom to
top) $q =$ 0.45, 1.3, 1.8, 2.5 $\times 10^{4}$ cm$^{-1}$.
\textbf{Inset:} The initial decay rate of the spin grating as a
function of $q$. Points are $\gamma_q -\tau_s^{-1}$; $\tau_s$ is
obtained from decay of homogenous ($q=0$) spin excitation. Error
bars are the size of the points except as shown. \textbf{Lower
points and line:} Room temperature. The line is a fit of the data to
$\gamma_q=\tau_s^{-1}+D_sq^2$, giving a spin diffusion length $L_s =
(D_s \tau_s)^{1/2} =$ 0.81 $\mu m$ and a ``spin mean-free-path'' $l
= 2D_s/v_F = 60$ nm. The observation of diffusive motion is
internally consistent, as $l$ is much smaller than both $L_s$ and
the smallest grating wavelength, 2.5 $\mu m$. \textbf{Upper points
and line:} 5K. The line has slope=1, corresponding to ballistic,
rather than diffusive, spin-motion with a velocity of 2.3$\times
10^7$ cm/s.} \label{fig:gammaVsT}
\end{figure}

We begin by discussing the decay rate where it varies slowly,
\textit{i.e.}, below 50 K and above 100 K. In the high-$T$ region
the spin dynamics can be accurately described in terms of
independent processes of spin diffusion and spin relaxation. In this
description, the decay rate varies with \textit{q} quadratically, as
$\gamma_q=\tau_s^{-1}+D_sq^2$, where $D_s$ is the spin diffusion
coefficient and $\tau_s$ is the spin relaxation
time\protect{\cite{Cameron}}. In the inset to Fig. 1 we plot
$\gamma_q-\tau_s^{-1}$ \textit{vs. q} at 295 K (lower points) and 5
K (upper points), on logarithmic axes. Here $1/\tau_s$ is
independently determined from the decay rate of the circular
dichroism induced by a circularly polarized pump
beam\protect{\cite{Bar-Ad}} (see supplementary information). A
comparison of the 295 K data with a line of slope two shows that the
decay of the grating is well described by diffusive dynamics, with
$D_s =$ 130 cm$^2$/s and $\tau_s$ = 50 ps.

Next, we examine the spin-grating dynamics at $T<$ 50 K. As shown in
the inset, the initial decay rates at 5 K are linear in \textit{q}
at the higher wavevectors. The change in power law exponent from two
to one indicates that a crossover from diffusive to ballistic
dynamics takes place as the sample temperature is lowered. In the
ballistic regime electrons propagate a distance comparable to the
grating wavelength, $\Lambda$, without scattering and the initial
decay rate is $\sim v_Fq$, the reciprocal of the time required for
an electron moving with the Fermi velocity to traverse a distance
$\Lambda/2\pi$.

Although the grating's initial decay rate saturates near $v_Fq$
when $T$ reaches $\sim$ 50 K, its time dependence continues to
change as $T$ is lowered further. Fig. 2 shows the grating
amplitude as function of time for several temperatures between 5 K
and 100 K, measured with a grating wavevector of 2.5$\times10^4$
cm$^{-1}$ (the $T$ indicated is the lattice temperature, which is
below the electron temperature, as will be discussed later). An
oscillatory structure appears in the decay curves, becoming
increasingly pronounced as $T$ decreases. The growth of these
oscillations is a consequence of the increase of the
mean-free-path, $l$, in the regime where $ql\geq1$.

\begin{figure}
\protect{\includegraphics[width=3in]{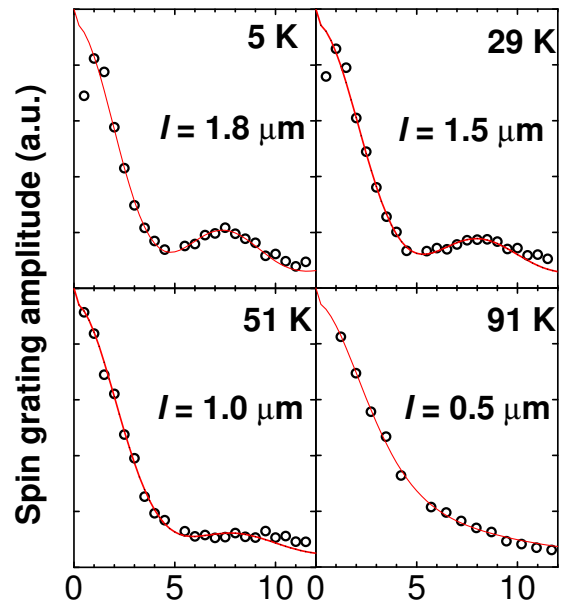}}
\caption{Time-dependence of the spin-grating's amplitude. The lines
are fits of the data to $S(q,\omega)$. The values of $l$ determined
from these fits are indicated in each panel. Due to laser heating,
the temperature $T_e$ of the electron gas is higher than the lattice
temperatures indicated.} \label{fig:TracesAndFits}
\end{figure}

To determine $D_s$ from data such as those in Fig. 2 we use an
expression for the time dependence of a spin fluctuation that is
applicable throughout the diffusive-ballistic crossover regime. If a
spin polarization wave is introduced at $t=0$, its subsequent
time-dependence is the Fourier transform of
$S(q,\omega)\propto[i\omega - D(q,\omega )q^2]^{-1}$, where
$D(q,\omega)$ is the dynamic spin diffusivity. In the limit $q \ll
k_F$,
\begin{equation} D(q,\omega ) = \frac{v_F/2}{\sqrt{(i\omega/v_F-1/l)^2 +
q^2}}, \label{eq:JMFforD}\end{equation} where Eq. \ref{eq:JMFforD}
extrapolates from the small-$q$ limit\protect{\cite{Rammer}} to the
ballistic regime. In attempting to fit the grating decay curves in
the plateau regime, we found that Eq. \ref{eq:JMFforD} is not quite
sufficient to describe the data. It is necessary to add to the
Fourier transform of $S(q,\omega)$ a small, slowly decaying
exponential with relative initial amplitude $\approx 0.1$ and
characteristic time $\approx$ 25 ps. We speculate that this slow
exponential may originate from a small fraction of localized
electrons. The solid lines through the data points in Fig. 2 show
the results of the fitting procedure, with fitting parameters $l$,
$v_F$, and the amplitude and time constant of the slow exponential.
Despite the complicating presence of the slow exponential, we
believe that the fits give an accurate indication of $l$, as this is
the only parameter that determines the rate at which the
oscillations are damped. Finally, the spin diffusion coefficient is
determined from the relation $D_s=v_Fl/2$.

The temperature dependence of $D_s$ obtained from our analysis of
the spin-grating dynamics is shown in Fig. 3 for QW's of different
electron density. For the two lower density samples (middle and
lower panels), the dynamics were diffusive at all $T$, consistent
with their lower mobility. To characterize charge transport in the
same set of samples, we performed 4-probe measurements of the 2D
charge conductance, $\sigma_c$, carrier density, $n$, and mobility,
$\mu$, on chips from the same set of wafers. Together, these
measurements allow us to test the assumption that the scattering
processes that control spin diffusion and charge conduction are the
same.  The link between conductance and diffusion coefficient is the
Einstein relation, $D_s=\sigma_s/e^2\chi_s$, where $\sigma_s$ and
$\chi_s$ are the spin conductance and susceptibility, respectively.
If the spin and charge scattering rates were the same (\textit{i.e.}
$\sigma_c=\sigma_s$), then $D_s$ would equal
$(\chi_0/\chi_s)D_{c0}$,\protect{\cite{CastellaniThermal}} where
$D_{c0}\equiv\sigma_c/e^2\chi_0$ and $\chi_0 = N_F(1-e^{-E_F/k_BT})$
is the noninteracting susceptibility (see supplementary information;
$N_F$ is the density of states at the Fermi energy and $k_B$ is
Boltzmann's constant). Physically, $D_{c0}$ is the quasiparticle
diffusion coefficient \protect{\cite{CastellaniThermal}},
approaching $\mu E_F/e$ and $\mu k_BT/e$ in the degenerate and
nondegenerate regimes, respectively. $D_{c0}$, calculated from the
4-probe transport data and plotted in Fig. 3, is considerably larger
than $D_s$ at all $T$ and for each of the samples. The ratio is far
greater than can be accounted for by many-body enhancement of the
spin susceptibility, as the factor $\chi_s/\chi_0$ is less than 1.4
in this range of electron
density\protect{\cite{YarlagaddaSpinSusceptibility, KwonQuantumMC}}.

\begin{figure}
\protect{\includegraphics[width=3.1in]{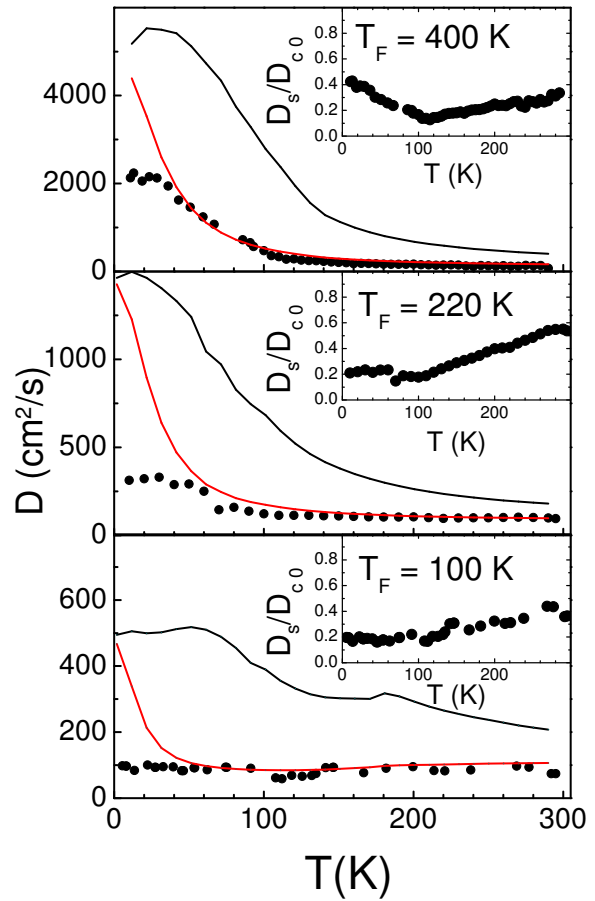}}
\caption{Comparison of motion of spin and charge, for samples with
the Fermi temperatures shown. \textbf{Dots (main panels):}
Spin-diffusion coefficients $D_s$ determined from optical
measurements. \textbf{Black lines:} Quasiparticle diffusion
coefficients $D_{c0}$ determined from transport data.
\textbf{Insets:} $D_s/D_{c0}$. \textbf{Red lines:} $D_s$ predicted
from spin Coulomb drag theory, taking $\chi_{s} = \chi_{0}$.}
\label{fig:DsAndInset}
\end{figure}

The contrast in the diffusion coefficients of charge and spin is
surprising, as the assumption $D_s = D_{c0}$ is widely used in
modeling spin transport in semiconductors. However, this assumption
fails to take into account \textit{e-e} collisions, whose rate can
be much faster than those of impurity or phonon scattering. The
\textit{e-e} scattering events can be ignored in the description of
charge transport because they conserve total momentum. However, they
can have a profound effect on spin transport, as illustrated in Fig.
4. For the collision depicted between electrons with opposite spin,
the charge current is conserved while the spin current reverses
direction.

\begin{figure}
\protect{\includegraphics[width=3.1in]{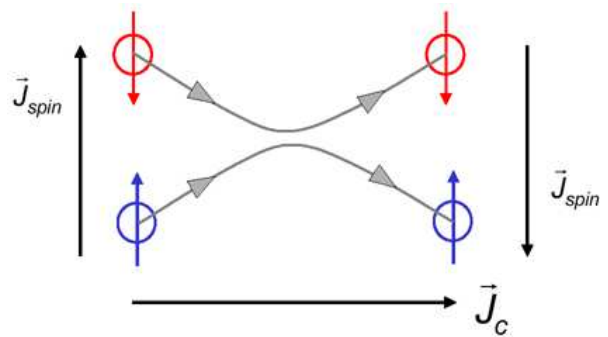}} \caption{A
representation of \textit{e-e} scattering that does not conserve
spin-current. Prior to the collision the spin-current is positive;
after, it is negative. The charge current does not change.}
\label{fig:Illustration}
\end{figure}

D'Amico and Vignale (DV) have proposed that the microscopic process
shown in Fig. 4 can change the nature of macroscopic spin transport.
Seen macroscopically, \textit{e-e} collisions transfer momentum
between the spin-up and spin-down populations, creating a force
damping their relative motion that DV term ``spin Coulomb drag''
(SCD)\protect{\cite{D'AmicoEurophysLett}}. Spin diffusion, which
requires a counterflow of the spin populations, is damped by SCD,
while charge diffusion is not. (The recently observed
\protect{\cite{KatoSpinHall, WunderlichSpinHall}} spin Hall effect
also involves the counterflow of spin populations, and so should be
damped by SCD.) According to DV
\protect{\cite{D'AmicoEurophysLett}}, the reduction of $D_s$
relative to $D_{c0}$ is:
\begin{equation} {D_s\over D_{c0}}=\left({\chi_0 \over \chi_s}\right){1\over {1+\mid\rho_{\uparrow\downarrow}\mid/\rho}},
\label{eq:DsAndRhoUpDown} \end{equation} where $\rho  = 1/\sigma_c$
is the charge resistance and $\rho_{\uparrow\downarrow}$ is the spin
drag resistance, parameterizing the rate of momentum exchange
between spin $\uparrow$ and $\downarrow$ electrons. DV, and
Flensberg and Jensen\protect{\cite{SpinDrag2dFlensberg}}, have
calculated $\rho_{\uparrow\downarrow}(T)$ for a 2DEG using the
random phase approximation (RPA), obtaining results that depend only
on the electron density of the quantum well.

Eq. \ref{eq:DsAndRhoUpDown} predicts that despite the complex $T$
dependences of the individual diffusion coefficients, their ratio
depends primarily on the single single factor,
$\mid\rho_{\uparrow\downarrow}\mid/\rho$.  We test this prediction
in Fig. 5, without invoking any assumptions or adjustable
parameters, by plotting $D_{c0}/D_s$ (the inverse of Eq. 2)
\textit{vs.} $\mid\rho_{\uparrow\downarrow}\mid/\rho$ for each of
the three samples measured in this study.  The transport
coefficients are taken directly from our measurements, while
$\mid\rho_{\uparrow\downarrow}\mid$ was calculated using Eq. 2 of
Ref. \protect{\cite{D'Amico2D_PRB}}. The resulting graph reveals the
simple linear dependence of $D_{c0}/D_s$ on
$\mid\rho_{\uparrow\downarrow}\mid/\rho$ predicted by Eq.
\ref{eq:DsAndRhoUpDown} over a large range of
$\mid\rho_{\uparrow\downarrow}\mid/\rho$, implying that SCD is
indeed the origin of the large suppression of $D_s$ relative to
$D_{c0}$. The fact that the slope is slightly greater than unity is
consistent with the expectation that the many-body enhancement of
$\chi_s$ relative to $\chi_0$ is small in this density
regime.\protect{\cite{YarlagaddaSpinSusceptibility, KwonQuantumMC}}
Finally, the fact that $D_{c0}/D_s$ extrapolates to near unity as
$\mid\rho_{\uparrow\downarrow}\mid/\rho\rightarrow0$ indicates that
the spin and charge diffusion coefficients approach each other in
the limit that the spin drag resistance becomes smaller than the
ordinary resistance. This result provides independent evidence that
the spin grating and four-probe techniques used in this work
accurately measure equilibrium spin and charge transport
coefficients, respectively.

Returning to the $T$-dependence shown in Fig. 3, the solid lines
show the prediction of Eq. \ref{eq:DsAndRhoUpDown} for $D_s$ with
the factor $\chi_0/\chi_s$ set equal to unity.  As could be
anticipated from the discussion of Fig. 5, SCD quantitatively
accounts for the suppression of $D_s$ relative to $D_{c0}$ over a
broad range of temperature and electron density.  It is clear,
however, that the measured $D_s$ consistently departs from theory
below 40 K. We believe that this discrepancy indicates that at low
$T$ the photoexcited electron gas does not cool to the lattice $T$.
If the electron gas retains the heat, $Q$, deposited by the
excitation, its temperature $T_e$ will rise to approximately
$(T^2+2Q/\beta)^{1/2}$, where $\beta=5.3\times10^5$ eV/cm$^2$-$K^2$
is the temperature coefficient of the electronic specific heat. We
estimate $Q=4\times10^8$ eV/cm$^2$, assuming that each absorbed
photon deposits approximately 10 meV (the energy width of the laser
pulse) into the Fermi sea. The resulting estimate for the minimum
$T_e$ is indeed $\sim$ 35 K.

Finally, we note that SCD can be highly advantageous for spintronic
applications, as it increases the distance that a spin packet can be
dragged by an electric field, $E$, before it spreads due to
diffusion.\protect{\cite{KikkawaLateralDrag}} The length $L_D$ that
a packet of width $w$ will drift before it broadens by a factor of
two is $w^2e\mu/D_s$. In the absence of SCD the ratio $\mu/D_s$
equals $e/k_BT_F$ or $e/k_BT$ in the degenerate or nondegenerate
regimes, respectively, and $L_D/w$ is independent of the underlying
scattering rates. In the degenerate regime, for example,
$L_D/w=eEw/kT_F$; drifting a spin packet farther than $w$ is only
possible in a strong $E$ limit, where the potential drop across the
packet exceeds the Fermi energy. Introducing SCD slows the
counterflow of spin $\uparrow$ and $\downarrow$ electrons without
affecting their co-propagation, amplifying $L_D/w$ by the factor
$1+|\rho_{\uparrow\downarrow}(T)|/\rho$. Clean materials with strong
\textit{e-e} scattering will have the largest values of
$\rho_{\uparrow\downarrow}(T)/\rho$, and hence be the best media for
propagation of spin information.

\begin{figure}
\protect{\includegraphics[width=3.1in]{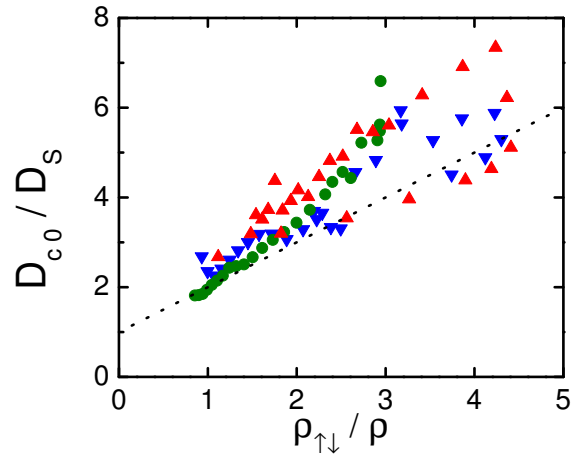}}
\caption{Relation between suppression of spin diffusion and spin
drag resistance. \textbf{X-axis:} ratio of
$\rho_{\uparrow\downarrow}$, determined from SCD
theory\protect{\cite{D'Amico2D_PRB}}, to measured resistivity
$\rho$. \textbf{Y-axis:} ratio of quasiparticle diffusion
coefficient, $D_{c0}$, to spin diffusion coefficient, $D_s$.
Temperature is an implicit parameter. Points are for samples with
$T_F$ = 400 K (\textbf{red}), 220 K (\textbf{green}), and 100 K
(\textbf{blue}). Points for $T <$ 40 K are not shown because the
electrons do not cool below 40 K. \textbf{Orange:} Points for a
sample with $T_F$ = 400 K, with $\rho$ increased by depositing a
portion of the Si dopants into the well. \textbf{Line:} has unity
slope and intercept, indicating the prediction of Eq.
\ref{eq:DsAndRhoUpDown} for $\chi_{s} = \chi_{0}$. For points above
the line, $\chi_s
> \chi_0$.} \label{fig:Collapsed data}
\end{figure}

\section{Quantum well characteristics.} The
GaAs/Ga$_{0.7}$Al$_{0.3}$As samples were grown in the (100)
direction by molecular beam epitaxy, and each consist of ten quantum
wells of thickness 12 nm, separated by 48 nm barriers. The Si
impurities were deposited in eight single atomic layers in the
center 14 nm of each barrier to maximize their distance from the
2DEG. The carrier concentration, \textit{n}, mobility, $\mu$, and
electrical resisitivity $\rho$ were measured using 4-probe transport
techniques without illumination. For the samples with \textit{n} of
7.8, 4.3, and 1.9 $\times 10^{11}$ cm$^{-2}$ per quantum well, at
low temperature $\mu$ reached 240,000, 92,000, and 69,000
cm$^2$/V-s, respecively.

\section{Optical methods.} The two interfering beams that generate
the optical-helicity wave derive from a Ti:Sapphire laser, which
produces a train of optical pulses with duration 100 fs, interpulse
separation 11 ns and center wavelength 820 nm. The incident power
density for most measurements was $\sim$ 500 W/cm$^{2}$,
corresponding to $\sim$ 6 W/cm$^{2}$ absorbed per quantum well. For
$T>$ 35 K grating decay rates did not change when measured at
incident powers down to 100 W/cm$^{2}$, suggesting that photoinduced
holes do not play a significant role in the electron spin transport
(typical electron-hole recombination times were $\sim$ 750 ps). At
low $T$ the grating decay rate increased slowly with decreasing
power, consistent with the electron heating model described in the
main text.

The grating wavevector was directed along the GaAs (01$\bar{1}$)
direction. To detect the induced spin grating, we mix diffracted and
transmitted probes to produce a photodetector current linear in the
diffracted field\protect{\cite{HeterodyneVohringer, HeterodyneChang,
GedikHetrodyne}}. The decisive advantage in this scheme is realized
by modulating the relative phase of these two beams sinusoidally at
1.2 kHz. Synchronous detection with a lock-in amplifier at the
modulation frequency leads to considerable rejection of laser noise
and stray light.

\bibliography{BibliographyComplete050823}

We thank I. D'Amico and G. Vignale for sending us numerical
evaluations of their integral expression for the spin drag
resistance. This work was funded by the US DOE, DARPA, and NSF-DMR.
We also gratefully acknowledge support from the Fannie and John
Hertz Foundation (C.P.W.) and the Hellman Foundation (J.E.M.).

Correspondence and requests for materials should be addressed to
C.P.W.~(email: cpweber@lbl.gov).

42.65.Hw, 72.25.b, 75.40.Gb.

\end{document}